\documentclass[12pt]{article}
 \usepackage{epsfig}
 \def\be{\begin{equation}}
 \def\ee{\end{equation}}
 \def\bea{\begin{eqnarray}}
 \def\eea{\end{eqnarray}}
 \usepackage{graphicx}

 \catcode`\@=11
 \def\lsim{\mathrel{\mathpalette\@versim<}}
 \def\gsim{\mathrel{\mathpalette\@versim>}}
 \def\@versim#1#2{\vcenter{\offinterlineskip
 \ialign{$\m@th#1\hfil##\hfil$\crcr#2\crcr\sim\crcr } }}
 \catcode`\@=12

 \catcode`@=12
\begin{document}
 \thispagestyle{empty}
 \begin{flushright}
 UCRHEP-T628\\
 Sep 2024\
 \end{flushright}
 \vspace{0.6in}
 \begin{center}
 {\LARGE \bf Light Dark Fermion in Two Natural Scenarios\\}
 \vspace{1.5in}
 {\bf Ernest Ma\\}
 \vspace{0.1in}
{\sl Department of Physics and Astronomy,\\ 
University of California, Riverside, California 92521, USA\\}
 \vspace{1.5in}

\end{center}

\begin{abstract}\
Dark matter is postulated as a light fermion in two natural scenarios 
as the outcome of a softly broken discrete symmetry. It is produced from 
the naturally suppressed decay of the standard-model Higgs boson through 
the freeze-in mechanism.
\end{abstract}

\newpage
\baselineskip 24pt
\noindent \underline{\it Introduction}~:~
The existence of dark matter is widely accepted, but its nature remains 
unknown.  It has eluded direct direction despite intensive experimental 
efforts to date.  Perhaps, it is a light fermion with feeble interactions with 
known matter, so that it is very difficult to observe.  In that case, the 
theoretical questions are why it is so light and why its interactions with the 
standard-model particles are so weak.  In this paper, it is shown how two 
simple extensions of the standard model allow these outcomes from an 
automatic softly broken discrete symmetry.

\noindent \underline{\it Gauged $B-L$ Symmetry}~:~
In the standard model, neutrinos are light fermions. They interact weakly 
with charged leptons.  It has long been known that they could not explain 
the observed cosmic structure attributed to dark matter.  On the other hand, 
if a light 
singlet fermion exists beyond the standard three neutrinos, it could be 
a candidate for dark matter.  From its mixing with the other 
neutrinos, it is itself unstable and decays into a neutrino and a 
photon.  This so-called sterile neutrino has long been studied as warm 
dark matter, and severe experimental constraints are obtained on its 
mass and couplings. 

To distinguish neutrinos from a singlet fermion, it is proposed that the 
standard model be extended to include the well-known gauged $B-L$ symmetry. 
Added to the usual particles are three right-handed singlet fermions 
$N_R$ with $B-L=-1$ and one singlet Higgs scalar $\chi^0$ with $B-L=2$.
This is a very well-known scenario for light seesaw neutrino masses, 
where $\langle \chi^0 \rangle \neq 0$ breaks $U(1)_{B-L}$ and allows 
$N_R$ to acquire a large Majorana mass, whereas $\nu_L$ pairs up with $N_R$ 
for a Dirac mass through $\langle \phi^0 \rangle \neq 0$ from the Higgs doublet 
$\Phi = (\phi^+,\phi^0)$ of the standard model.
  
\noindent \underline{\it $Z_4 \to Z_2$ Symmetry}~:~
To accommodate dark matter, it is proposed that one singlet fermion $S_L$ 
and one singlet real scalar $\sigma$ are added without any transformation 
under any symmetry of the $B-L$ model.  Nevertheless, in any renormalizable 
Lagrangian before spontaneous symmetry breaking, there is always a discrete 
$Z_4$ symmetry~\cite{m23} under which
\begin{equation}
A_\mu \sim 1, ~~~ \phi \sim -1, ~~~ \psi_L \sim i, ~~~ \psi_R \sim -i, 
\end{equation}
where $A_\mu$ is any vector gauge field, $\phi$ any scalar, $\psi_L$ any 
left-handed chiral fermion, and $\psi_R$ any right-handed fermion.  This is 
always the case without the soft trilinear terms (i.e. $\phi^3$, 
$\psi_L\psi_L$, $\psi_R \psi_R$, $\bar{\psi}_L \psi_R$) which obviously would 
break $Z_4$ explicitly to $Z_2$, which is then just the ordinary trivial 
symmetry of fermions versus bosons.  In the standard model before symmetry 
breaking, the Lagrangian has no trilinear terms.  Hence $Z_4$ is simply a 
subgroup of the imposed symmetries of the standard model, and is not useful 
for gaining any new insight into the underlying physics. 

In the gauged $B-L$ model extended by $S_L$ and $\sigma$, the situation 
changes and $Z_4 \to Z_2$ becomes a useful guide to understanding 
freeze-in light dark matter.
\begin{table}[tbh]
\centering
\begin{tabular}{|c|c|c|}
\hline
particle/interaction & $B-L$ & $Z_4$ \\ 
\hline
$\nu_L$ & $-1$ & $i$ \\ 
$N_R$ & $-1$ & $-i$ \\
$\phi^0$ & 0 & $-1$ \\ 
$\chi^0$ & 2 & $-1$ \\ 
\hline 
$S_L$ & 0 & $i$ \\ 
$\sigma$ & 0 & $-1$ \\ 
\hline
\hline
$\bar{N}_R \nu_L \phi^0$ & yes & yes \\ 
$N_R N_R \chi^0$ & yes & yes \\ 
\hline
$S_L \nu_L \phi^0$ & no & yes \\ 
$\bar{S}_L N_R \chi^0$ & no & yes \\ 
$\bar{S}_L N_R \sigma$ & no & yes \\ 
\hline
\end{tabular}
\caption{Particles and interactions of extended $B-L$ model}
\end{table}

Reading from Table 1, it is clear that $Z_4$ is not a subgroup of the 
symmetries of the $B-L$ model.  If only the standard model were considered, 
$\bar{S}_L$ would be identical to $N_R$ and $Z_4$ would be derivable from 
assigning lepton number $L=1$ to both, in which case $Z_4 = (i)^{B-L+4Y}$, 
where $Q=I_3+Y$. 
As it is, the explicit breaking of $Z_4$ by soft trilinear terms is a 
natural framework for freeze-in light dark matter, as explained below.

\noindent \underline{\it Light Fermion Dark Matter}~:~ 
Terms in the Lagrangian involving $S$ are 
\begin{equation}
{\cal L}_S = {1 \over 2} m_2 S S + {1 \over 2} f \sigma S S + H.c.
\end{equation}
This shows that the $f$ coupling obeys $Z_4$, but $m_2$ breaks it and 
may thus be assumed naturally small. 
However, $S$ is stable because of the residual $Z_2$ symmetry. 
The $B-L$ breaking scale is assumed much larger than the electroweak scale. 
Hence the scalar $\chi$ field may be integrated away, resulting in the 
Higgs potential involving $\Phi$ and $\sigma$ as 
\begin{eqnarray}
V &=& -\mu_0^2 \Phi^\dagger \Phi + {1 \over 2} \lambda_0 
(\Phi^\dagger \Phi)^2 + {1 \over 2} m_1^2 \sigma^2 + 
{1 \over 6} \mu_1 \sigma^3 + {1 \over 24} \lambda_1 \sigma^4 \nonumber \\ 
&+& \mu_2 \sigma \Phi^\dagger \Phi + {1 \over 2} \lambda_{2} \sigma^2 
\Phi^\dagger \Phi, 
\end{eqnarray}
where the terms $\mu_1, \mu_2$ break $Z_4$ to $Z_2$, and may be assumed 
naturally small.

In the parameter space $m_1^2 > \mu_0^2 > 0$, both $\phi^0$ and $\sigma$ 
obtain vacuum expectation values:
\begin{equation}
\langle \phi^0 \rangle = v_0 \simeq \sqrt{\mu_0^2 \over \lambda_0},
\end{equation}
and~\cite{m01}
\begin{equation}
\langle \sigma \rangle = v_1 \simeq {-\mu_2 v_0^2 \over m_1^2 + 
\lambda_{2} v_0^2},
\end{equation}
which is naturally small as it is proportional to  
$\mu_2$ which breaks $Z_4$.  The mass of $S$ is now 
\begin{equation}
m_S = f v_1 + m_2,
\end{equation}
which remains small.  There are two physical scalars after spontaneous 
symmetry breaking, i.e. the standard Higgs boson 
$h$ and the new $\sigma$ shifted by $v_1$.  They have masses given by 
\begin{equation} 
m_h^2 = 2 \lambda_0 v_0^2, ~~~ m_\sigma^2 \simeq m_1^2 + \lambda_{2} v_0^2, 
\end{equation}
and their mixing is naturally small, i.e.
\begin{equation}
\theta_{h\sigma} \simeq {\sqrt{2} \mu_2 v_0 \over m_1^2} \simeq 
{-\sqrt{2} v_1 \over v_0},
\end{equation} 
assuming that $m_\sigma^2 \simeq m_1^2 >> m_h^2$.  If $m_2$ in Eq.~(6) is also 
neglected, then the coupling of $h$ to $SS$ is  
\begin{equation}
f_h = f \theta_{h\sigma} \simeq {-\sqrt{2} m_S \over v_0},
\end{equation}
and the decay rate of $h \to SS + \bar{S}\bar{S}$ is
\begin{equation}
\Gamma_h = {f_h^2 m_h \over 8 \pi} \sqrt{1-4r^2} (1-2r^2),
\end{equation}
where $r=m_S/m_h$.  Now $S$ may be produced by $h$ in a freeze-in scenario 
with the correct relic abundance if~\cite{ac13} 
\begin{equation}
f_h \sim 10^{-12} r^{-1/2}.
\end{equation}
Matching the two $f_h$ expressions, $m_S \sim 1.2$ keV is obtained. 
Note that this model may 
be expanded to include any number of $S$ and $\sigma$ particles.  There is 
always a lightest fermion singlet which is a natural candidate for dark 
matter.

\noindent \underline{\it Dark $U(1)_D$ Model}~:~
Another scenario for the $Z_4 \to Z_2$ application is based on 
the original very well-known spontaneously broken Abelian $U(1)$ gauge 
model~\cite{h64}.  It begins with a vector gauge boson (call it $Z_D$) 
and a complex scalar (call it $\zeta$).  It is 
kwown~\cite{fr12,bkps13,dft16,dm17,m17} to have 
automatic charge conjugation invariance, i.e. $Z_D \to -Z_D$, 
$\zeta \to \zeta^*$, resulting in $g_D \to -g_D$.  After spontaneous 
symmetry breaking, the above still holds, i.e. $Z_D \to -Z_D$, 
$\zeta_R \to \zeta_R$, and $\zeta_I \to -\zeta_I$ which becomes the 
longitudinal component of the now massive $Z_D$.  This fact 
has been used to suggest that $Z_D$ may be dark matter. 

Since the standard model (SM) of quarks and leptons has a gauge $U(1)_Y$ 
factor, the addition of gauge $U(1)_D$ allows for the gauge-invariant 
kinetic mixing~\cite{h86} of the two associated gauge bosons, so $Z_D$ may 
mix with the $U(1)_Y$ gauge boson, of which the photon is a component.  
This special but rather peculiar possibility has spawned many studies of a 
so-called light dark photon, and the experiments which may be relevant in 
finding it~\cite{dark16}.  On the other hand, this arbitrary kinetic mixing 
term may be forbidden by invoking the dark charge conjugation symmetry, as 
is assumed in the following.

In the Higgs model, $Z_D$ is the sole dark matter.  In Ref.~\cite{m17}, 
a Dirac fermion (call it $\psi$) is added, transforming also under $U(1)_D$, 
then the Lagrangian is also invariant under dark charge conjugation, as 
well as the global $U(1)$ transformation operating on $\psi$, i.e. 
dark fermion number.  Hence $\psi$ is a dark-matter candidate. 
The decay $Z_D$ to $\bar{\psi} \psi$ is allowed by charge conjugation 
symmetry, but is forbidden in Ref.~\cite{m17} by the choice 
$m_\psi >> m_{Z_D}$.  However, the Dirac mass term breaks $Z_4$, so 
the choice should be $m_\psi << m_{Z_D}$ instead.  This changes 
the structure of the model completely and results in a simple model of 
light dark fermion from Higgs decay~\cite{m19} in a freeze-in 
scenario~\cite{mmy93,ckkr01,m02,aim07,hjmw10}, as shown below.

\noindent \underline{\it Light Fermion Dark Matter}~:~
The model Lagrangian is the same as that of Ref.~\cite{m17} except for the 
hierarchical choice of parameters.  It assumes $U(1)_D$ gauge symmetry, 
requiring thus a vector gauge boson $Z_D$.  It is spontaneously broken by 
a complex scalar $\zeta$ with charge $g_D$.  A Dirac fermion $\psi$ is also 
present with charge $g_\psi$.  The complete Lagrangian before symmetry 
breaking is
\begin{eqnarray}
{\cal L} &=& -{1 \over 4} (\partial^\mu Z_D^\nu - \partial^\nu Z_D^\mu) 
(\partial_\mu Z_{D\nu} - \partial_\nu Z_{D\mu}) + (\partial^\mu \zeta - 
ig_D Z_D^\mu \zeta)(\partial_\mu \zeta^* + i g_D Z_{D\mu} \zeta^*) 
\nonumber \\ 
&+& \mu_D^2 \zeta^* \zeta - {1 \over 2} \lambda_D (\zeta^* \zeta)^2 
+ i \bar{\psi} \gamma_\mu (\partial^\mu - ig_\psi Z_D^\mu) \psi 
- m_\psi \bar{\psi} \psi.
\end{eqnarray}
In the above, if $g_D$ is replaced by $-g_D$, $\zeta$ by $\zeta^*$, 
$g_\psi$ by $-g_\psi$, and $\psi$ by its dark charge conjugate, 
exactly the same physical theory is obtained.  Armed with the knowledge of 
Ref.~\cite{m23}, it is also clear that ${\cal L}$ respects the universal 
$Z_4$ symmetry mentioned already, i.e.
\begin{equation}
Z_D \sim 1, ~~~ \zeta \sim -1, ~~~ \psi_L \sim i, ~~~ \psi_R \sim -i,
\end{equation}
except for the soft dimension-three $m_\psi$ term.  This means that 
$m_\psi$ should be chosen much smaller than $\mu_D$, instead of the 
other way round~\cite{m17}.

The spontaneous breaking of $U(1)_D$ with 
$\langle \zeta \rangle = v_D/\sqrt{2}$ changes the Lagrangian to
\begin{eqnarray}
{\cal L} &=& -{1 \over 4} (\partial^\mu Z_D^\nu - \partial^\nu Z_D^\mu) 
(\partial_\mu Z_{D\nu} - \partial_\nu Z_{D\mu}) + {1 \over 2} m^2_{Z_D} 
Z_D^\mu Z_{D\mu} + {1 \over 2}(\partial^\mu h_D)(\partial_\mu h_D) 
- {1 \over 2} m^2_{h_D} h_D^2 \nonumber \\ &+& {m^2_{h_D} \over 2 v_D} h_D^3 + 
{m^2_{h_D} \over 8 v_D^2} h_D^4 + g_D^2 v_D h_D (Z^\mu_D Z_{D\mu}) 
+ {1 \over 2} g_D^2 h_D^2 (Z^\mu_D Z_{D\mu}) \nonumber \\ &+& 
i \bar{\psi} \gamma_\mu \partial^\mu \psi - m_\psi \bar{\psi} \psi + 
g_\psi Z^\mu_D \bar{\psi} \gamma_\mu \psi,
\end{eqnarray}
where $v_D^2 = 2 \mu_D^2/\lambda_D$, $h_D = \sqrt{2} Re(\zeta)-v_D$, 
$m_{Z_D} = g_D v_D$, and $m^2_{h_D} = \lambda_D v_D^2$.  
The hierarchy $m_\psi << m_{Z_D}$ is assumed.  Note that $g_\psi$ is 
independent of $g_D$.

Three new particles exist beyond those of the SM, i.e. $Z_D$, $h_D$, and 
$\psi$.   The light dark fermion $\psi$ is a Dirac particle with a 
conserved dark fermion number.  It couples to $Z_D$ directly, but 
only indirectly to $h_D$ through $Z_D$. 
Assuming that the reheat temperature of the Universe is well below 
$m_{Z_D}$ and $m_{h_D}$, then $\psi$ is not thermally produced. 
However, since $h_D$ must mix with the SM Higgs $h$, the latter's 
decay to $\bar{\psi}\psi$ could account for the correct relic abundance 
of dark matter in the Universe.

Consider the extended scalar potential involving both $\zeta$ and the 
SM Higgs doublet $\Phi = (\phi^+,\phi^0)$:
\begin{eqnarray}
V &=& -\mu_D^2 \zeta^* \zeta + {1 \over 2} \lambda_D (\zeta^* \zeta)^2 
- \mu_h^2 \Phi^\dagger \Phi + {1 \over 2} \lambda_h (\Phi^\dagger \Phi)^2 
+ \lambda_{hD} (\zeta^* \zeta)(\Phi^\dagger \Phi).
\end{eqnarray}
Using $\phi^0 = (v_h+h)/\sqrt{2}$, the $2 \times 2$ mass-squared 
matrix spanning $(h_D,h)$ is given by
\begin{equation}
{\cal M}^2_{h_D,h} = \pmatrix{\lambda_D v_D^2 & \lambda_{hD} v_D v_h \cr 
\lambda_{hD} v_D v_h & \lambda_h v_h^2}.
\end{equation}
Assuming $m_{h_D} >> m_{h} = 125$ GeV, the $h_D-h$ mixing is then 
\begin{equation}
\theta_{hD} \simeq {\lambda_{hD} v_h \over \lambda_D v_D}.  
\end{equation}

\noindent \underline{\it Higgs decay to $\bar{\psi}\psi$}~:~
Through $\theta_{hD}$, the SM Higgs boson $h$ may decay to $\bar{\psi}\psi$ 
in one loop as shown in Fig.~1. 
\begin{figure}[htb]
 \vspace*{-7cm}
 \hspace*{-3cm}
 \includegraphics[scale=1.0]{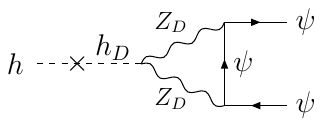}
 \vspace*{-22.0cm}
 \caption{Decay of $h$ to $\bar{\psi}\psi$.}
 \end{figure}
The effective coupling is
\begin{equation}
f_h = {g_D^2 g_\psi^2 \over 2 \pi^2} {\lambda_{hD} \over \lambda_D} 
{v_h m_\psi \over m^2_{Z_D}}.
\end{equation}
Let $m_{Z_D} = 1$ TeV, $\lambda_{hD}/\lambda_D = 0.1$, $g_D = g_\psi = e$, 
then Eq.~(11) is satisfied with $m_\psi = 10$ MeV.

\noindent \underline{\it Concluding Remarks}~:~
If dark matter is a light fermion, two natural scenarios have been presented 
where an automatic softly broken discrete symmetry allows it to be light 
and its interactions with the standard model are suppressed.  Both scenarios 
are gauge $U(1)$ extendsions of the standard model, the first $B-L$ and the 
second dark $U(1)_D$.  The light dark singlet fermion is produced through 
the freeze-in mechanism from the deccay the SM Higgs boson.

\noindent \underline{\it Acknowledgement}~:~
This work was supported in part by the U.~S.~Department of Energy Grant 
No. DE-SC0008541.

\bibliographystyle{unsrt}

\end{document}